\documentclass{emulateapj}

\usepackage{graphicx}
\usepackage{amsmath,amstext}
\usepackage{xcolor}
\usepackage{lineno}
\usepackage{hyperref}
\usepackage[figure,figure*]{hypcap}

\makeatletter
\renewcommand*\frontmatter@preabstractspace{2in\relax}
\makeatother

\newcommand{\samplesize}{214}
\newcommand{\NLowz}{126}
\newcommand{\samplename}{DES-SN3YR}
\newcommand{\wcorr}{-0.002}
\newcommand{\wcorrerr}{0.008}
\newcommand{\omcorr}{0.000}
\newcommand{\omcorrerr}{0.002}

\newcommand{\simwcorr}{0.0005}
\newcommand{\fiftysimwcorrmean}{0.007}
\newcommand{\fiftysimwcorrerr}{0.008}

\newcommand{\simomcorr}{0.000}
\newcommand{\fiftysimomcorrmean}{0.001}
\newcommand{\fiftysimomcorrerr}{0.001}

\newcommand{\dmu}{\Delta \mu}
\newcommand{\dax}{\Delta \alpha x_1}
\newcommand{\dbc}{\Delta \beta c}
\newcommand{\dmB}{\Delta m_B}
\newcommand{\dw}{\Delta w}
\newcommand{\dOm}{\Delta \Omega_m}
\newcommand{\Fsrc}{F_{\nu, \rm src}}
\newcommand{\Fref}{F_{\nu, \rm ref}}
\newcommand{\lowz}{low-$z$}
\newcommand{\Phiatmobs}{\phi^{\rm atm}_{\rm obs}(\lambda)}
\newcommand{\Phiatmref}{\phi^{\rm atm}_{\rm ref}(\lambda)}

\newcommand{\Phitot}{\phi_{\rm b,tot}(\lambda)}
\newcommand{\Phiinstobs}{\phi^{\rm inst}_{ \rm b, obs}(\lambda)}
\newcommand{\Phiinstref}{\phi^{\rm inst}_{\rm b, ref}(\lambda)}
\newcommand{\SNIa}{SN\,Ia}
\newcommand{\SNeIa}{SNe\,Ia}
\newcommand{\angstrom}{\mbox{\normalfont\AA}}

\begin{document}
{
\vspace*{-\headsep}\vspace*{\headheight}
\footnotesize \hfill FERMILAB-PUB-18-443-PPD\\
\vspace*{-\headsep}\vspace*{\headheight}
\footnotesize \hfill DES-2017-0291
}
\title{FIRST COSMOLOGY RESULTS USING TYPE Ia SUPERNOVAE FROM THE DARK ENERGY SURVEY: EFFECTS OF CHROMATIC CORRECTIONS TO SUPERNOVA PHOTOMETRY ON MEASUREMENTS OF COSMOLOGICAL PARAMETERS}

\def\andname{}

\author{
J.~Lasker\altaffilmark{1,2},
R.~Kessler\altaffilmark{1,2},
D.~Scolnic\altaffilmark{2},
D.~Brout\altaffilmark{3},
C.~B.~D'Andrea\altaffilmark{3},
T.~M.~Davis\altaffilmark{4},
S.~R.~Hinton\altaffilmark{4},
A.~G.~Kim\altaffilmark{5},
C.~Lidman\altaffilmark{6},
E.~Macaulay\altaffilmark{7},
A.~M\"oller\altaffilmark{8,6},
M.~Sako\altaffilmark{3},
M.~Smith\altaffilmark{9},
M.~Sullivan\altaffilmark{9},
J.~Asorey\altaffilmark{10},
B.~A.~Bassett\altaffilmark{11,12},
D.~L.~Burke\altaffilmark{13,14},
J.~Calcino\altaffilmark{4},
D.~Carollo\altaffilmark{15},
M.~Childress\altaffilmark{9},
J.~Frieman\altaffilmark{16,2},
J.~K.~Hoormann\altaffilmark{4},
E.~Kasai\altaffilmark{17,12},
T.~S.~Li\altaffilmark{16,2},
M.~March\altaffilmark{3},
E.~Morganson\altaffilmark{18},
E.~S.~Rykoff\altaffilmark{13,14},
E.~Swann\altaffilmark{7},
B.~E.~Tucker\altaffilmark{8,6},
W.~Wester\altaffilmark{16},
T.~M.~C.~Abbott\altaffilmark{19},
S.~Allam\altaffilmark{16},
J.~Annis\altaffilmark{16},
S.~Avila\altaffilmark{7},
K.~Bechtol\altaffilmark{20},
E.~Bertin\altaffilmark{21,22},
D.~Brooks\altaffilmark{23},
A.~Carnero~Rosell\altaffilmark{24,25},
M.~Carrasco~Kind\altaffilmark{26,18},
J.~Carretero\altaffilmark{27},
F.~J.~Castander\altaffilmark{28,29},
L.~N.~da Costa\altaffilmark{25,30},
C.~Davis\altaffilmark{13},
J.~De~Vicente\altaffilmark{24},
H.~T.~Diehl\altaffilmark{16},
P.~Doel\altaffilmark{23},
A.~Drlica-Wagner\altaffilmark{16,2},
B.~Flaugher\altaffilmark{16},
J.~Garc\'ia-Bellido\altaffilmark{31},
E.~Gaztanaga\altaffilmark{28,29},
D.~Gruen\altaffilmark{13,14},
R.~A.~Gruendl\altaffilmark{26,18},
J.~Gschwend\altaffilmark{25,30},
G.~Gutierrez\altaffilmark{16},
D.~L.~Hollowood\altaffilmark{32},
K.~Honscheid\altaffilmark{33,34},
D.~J.~James\altaffilmark{35},
S.~Kent\altaffilmark{16,2},
E.~Krause\altaffilmark{36},
R.~Kron\altaffilmark{16,2},
K.~Kuehn\altaffilmark{37},
N.~Kuropatkin\altaffilmark{16},
M.~Lima\altaffilmark{38,25},
M.~A.~G.~Maia\altaffilmark{25,30},
J.~L.~Marshall\altaffilmark{39},
P.~Martini\altaffilmark{33,40},
F.~Menanteau\altaffilmark{26,18},
C.~J.~Miller\altaffilmark{41,42},
R.~Miquel\altaffilmark{43,27},
A.~A.~Plazas\altaffilmark{44},
E.~Sanchez\altaffilmark{24},
V.~Scarpine\altaffilmark{16},
I.~Sevilla-Noarbe\altaffilmark{24},
R.~C.~Smith\altaffilmark{19},
M.~Soares-Santos\altaffilmark{45},
F.~Sobreira\altaffilmark{46,25},
E.~Suchyta\altaffilmark{47},
M.~E.~C.~Swanson\altaffilmark{18},
G.~Tarle\altaffilmark{42},
D.~L.~Tucker\altaffilmark{16},
A.~R.~Walker\altaffilmark{19}
\\ \vspace{0.2cm} (DES Collaboration) \\
}

\affil{$^{1}$ Department of Astronomy and Astrophysics, University of Chicago, Chicago, IL 60637, USA}
\affil{$^{2}$ Kavli Institute for Cosmological Physics, University of Chicago, Chicago, IL 60637, USA}
\affil{$^{3}$ Department of Physics and Astronomy, University of Pennsylvania, Philadelphia, PA 19104, USA}
\affil{$^{4}$ School of Mathematics and Physics, University of Queensland,  Brisbane, QLD 4072, Australia}
\affil{$^{5}$ Lawrence Berkeley National Laboratory, 1 Cyclotron Road, Berkeley, CA 94720, USA}
\affil{$^{6}$ The Research School of Astronomy and Astrophysics, Australian National University, ACT 2601, Australia}
\affil{$^{7}$ Institute of Cosmology and Gravitation, University of Portsmouth, Portsmouth, PO1 3FX, UK}
\affil{$^{8}$ ARC Centre of Excellence for All-sky Astrophysics (CAASTRO)}
\affil{$^{9}$ School of Physics and Astronomy, University of Southampton,  Southampton, SO17 1BJ, UK}
\affil{$^{10}$ Korea Astronomy and Space Science Institute, Yuseong-gu, Daejeon, 305-348, Korea}
\affil{$^{11}$ African Institute for Mathematical Sciences, 6 Melrose Road, Muizenberg, 7945, South Africa}
\affil{$^{12}$ South African Astronomical Observatory, P.O.Box 9, Observatory 7935, South Africa}
\affil{$^{13}$ Kavli Institute for Particle Astrophysics \& Cosmology, P. O. Box 2450, Stanford University, Stanford, CA 94305, USA}
\affil{$^{14}$ SLAC National Accelerator Laboratory, Menlo Park, CA 94025, USA}
\affil{$^{15}$ INAF, Astrophysical Observatory of Turin, I-10025 Pino Torinese, Italy}
\affil{$^{16}$ Fermi National Accelerator Laboratory, P. O. Box 500, Batavia, IL 60510, USA}
\affil{$^{17}$ Department of Physics, University of Namibia, 340 Mandume Ndemufayo Avenue, Pionierspark, Windhoek, Namibia}
\affil{$^{18}$ National Center for Supercomputing Applications, 1205 West Clark St., Urbana, IL 61801, USA}
\affil{$^{19}$ Cerro Tololo Inter-American Observatory, National Optical Astronomy Observatory, Casilla 603, La Serena, Chile}
\affil{$^{20}$ LSST, 933 North Cherry Avenue, Tucson, AZ 85721, USA}
\affil{$^{21}$ CNRS, UMR 7095, Institut d'Astrophysique de Paris, F-75014, Paris, France}
\affil{$^{22}$ Sorbonne Universit\'es, UPMC Univ Paris 06, UMR 7095, Institut d'Astrophysique de Paris, F-75014, Paris, France}
\affil{$^{23}$ Department of Physics \& Astronomy, University College London, Gower Street, London, WC1E 6BT, UK}
\affil{$^{24}$ Centro de Investigaciones Energ\'eticas, Medioambientales y Tecnol\'ogicas (CIEMAT), Madrid, Spain}
\affil{$^{25}$ Laborat\'orio Interinstitucional de e-Astronomia - LIneA, Rua Gal. Jos\'e Cristino 77, Rio de Janeiro, RJ - 20921-400, Brazil}
\affil{$^{26}$ Department of Astronomy, University of Illinois at Urbana-Champaign, 1002 W. Green Street, Urbana, IL 61801, USA}
\affil{$^{27}$ Institut de F\'{\i}sica d'Altes Energies (IFAE), The Barcelona Institute of Science and Technology, Campus UAB, 08193 Bellaterra (Barcelona) Spain}
\affil{$^{28}$ Institut d'Estudis Espacials de Catalunya (IEEC), 08034 Barcelona, Spain}
\affil{$^{29}$ Institute of Space Sciences (ICE, CSIC),  Campus UAB, Carrer de Can Magrans, s/n,  08193 Barcelona, Spain}
\affil{$^{30}$ Observat\'orio Nacional, Rua Gal. Jos\'e Cristino 77, Rio de Janeiro, RJ - 20921-400, Brazil}
\affil{$^{31}$ Instituto de Fisica Teorica UAM/CSIC, Universidad Autonoma de Madrid, 28049 Madrid, Spain}
\affil{$^{32}$ Santa Cruz Institute for Particle Physics, Santa Cruz, CA 95064, USA}
\affil{$^{33}$ Center for Cosmology and Astro-Particle Physics, The Ohio State University, Columbus, OH 43210, USA}
\affil{$^{34}$ Department of Physics, The Ohio State University, Columbus, OH 43210, USA}
\affil{$^{35}$ Harvard-Smithsonian Center for Astrophysics, Cambridge, MA 02138, USA}
\affil{$^{36}$ Department of Astronomy/Steward Observatory, 933 North Cherry Avenue, Tucson, AZ 85721-0065, USA}
\affil{$^{37}$ Australian Astronomical Optics, Macquarie University, North Ryde, NSW 2113, Australia}
\affil{$^{38}$ Departamento de F\'isica Matem\'atica, Instituto de F\'isica, Universidade de S\~ao Paulo, CP 66318, S\~ao Paulo, SP, 05314-970, Brazil}
\affil{$^{39}$ George P. and Cynthia Woods Mitchell Institute for Fundamental Physics and Astronomy, and Department of Physics and Astronomy, Texas A\&M University, College Station, TX 77843,  USA}
\affil{$^{40}$ Department of Astronomy, The Ohio State University, Columbus, OH 43210, USA}
\affil{$^{41}$ Department of Astronomy, University of Michigan, Ann Arbor, MI 48109, USA}
\affil{$^{42}$ Department of Physics, University of Michigan, Ann Arbor, MI 48109, USA}
\affil{$^{43}$ Instituci\'o Catalana de Recerca i Estudis Avan\c{c}ats, E-08010 Barcelona, Spain}
\affil{$^{44}$ Jet Propulsion Laboratory, California Institute of Technology, 4800 Oak Grove Dr., Pasadena, CA 91109, USA}
\affil{$^{45}$ Brandeis University, Physics Department, 415 South Street, Waltham MA 02453}
\affil{$^{46}$ Instituto de F\'isica Gleb Wataghin, Universidade Estadual de Campinas, 13083-859, Campinas, SP, Brazil}
\affil{$^{47}$ Computer Science and Mathematics Division, Oak Ridge National Laboratory, Oak Ridge, TN 37831}

\begin{abstract}
Calibration uncertainties have been the leading systematic uncertainty in recent analyses using type Ia Supernovae (\SNeIa{}) to measure cosmological parameters.  To improve the calibration, we present the application of Spectral Energy Distribution (SED)-dependent ``chromatic corrections"  to the supernova light-curve photometry from the Dark Energy Survey (DES). These corrections depend on the combined atmospheric and instrumental transmission function for each exposure, and they affect photometry at the 0.01 mag (1\%) level, comparable to systematic uncertainties in calibration and photometry. Fitting our combined DES and \lowz{} \SNIa{} sample with Baryon Acoustic Oscillation (BAO) and Cosmic Microwave Background (CMB) priors for the cosmological parameters $\Omega_{\rm m}$ (the fraction of the critical density of the universe comprised of matter) and $w$ (the dark energy equation of state parameter), we compare those parameters before and after applying the corrections. We find the change in $w$ and $\Omega_{\rm m}$ due to not including chromatic corrections are $\wcorr{}$ and $\omcorr{}$, respectively, for the \samplename{} sample with BAO and CMB priors, consistent with a larger \samplename{}-like simulation, which has a $w$-change of $\simwcorr{}$ with an uncertainty of $\wcorrerr{}$ and an $\Omega_{\rm m}$ change of $\simomcorr{}$ with an uncertainty of $\omcorrerr{}$ . However, when considering samples on individual CCDs we find large redshift-dependent biases ($\sim 0.02$ in distance modulus) for supernova distances.
\end{abstract}



\section{introduction}
\label{intro}
Supernova cosmologists uses type Ia Supernovae (\SNeIa{}) as standardizable candles to measure distances over a wide range of redshifts, which, when combined with a measurement of the redshift, are used to trace the expansion history of the Universe. The SN Ia distances and redshifts are fit to a model that is typically parameterized in terms of the fraction of the universe's energy that is in matter ($\Omega_{\rm M}$) versus that which is in dark energy ($\Omega_{\Lambda}$), as well as the equation of state parameter of dark energy, $w$.

The recovery of cosmological parameters from SNe is sensitive to calibration in two ways.  First, cosmological constraints depend on  comparing the relative brightnesses of SNe at different redshifts. As the rest frame SN spectrum is redshifted, we observe it in different bandpasses which must be calibrated relative to each other. Second, we observe SNe at different positions on the sky, different locations on our focal plane, and in different weather conditions.  Non-uniformity of these observations can introduce potential cosmological biases. Together, these calibration uncertainties make up the largest source of systematic uncertainty on cosmological parameters derived from \SNIa{} distances.

The impact of the systematic uncertainty from calibration is well illustrated in the recent analysis of the Pantheon sample \citep{Pantheon}, which is the largest combined sample of spectroscopically confirmed \SNeIa{} analyzed to date. For the dark energy equation of state parameter $w$, the Pantheon analysis' calibration uncertainty of 2-6 mmag, depending on sample, contributes $\sigma_w = 0.02$, half of their total uncertainty on $w$. 

The samples included in this analysis are from the Pan-STARRS~1 (PS1, \citealt{R14}) Medium Deep Survey, the Sloan Digital Sky Survey~II (SDSS-II, \citealt{S14}), the Supernova Legacy Survey (SNLS, \citealt{SNLS}), HST \citep{R04, R07,CANDELS}, the Center for Astrophysics low redshift surveys (CFA3 and CFA4, \citealt{CfA3} and \citealt{CfA4}), the Carnegie Supernova Project (CSP, \citealt{CSP}), and the HST Cluster Supernova Survey \citep{Suzuki}. 

It's critical to note that while the calibration uncertainties from these samples are a factor of 50 below the distance uncertainties, the binned distance uncertainties that constrain cosmology are reduced as ($\sqrt{N_{\rm SN}}$), unlike the calibration error.

It is important to reduce calibration uncertainties in order to utilize the improved statistical power in measuring cosmological parameters from surveys with larger samples.  The Dark Energy Survey Supernova Program (DES-SN, \citealt{Diffimg}) is measuring multi-band light curves of a photometric sample of thousands of \SNeIa{}, as well as a spectroscopically classified sample of several hundred \SNeIa. Furthermore, the Large Synoptic Survey Telescope (LSST, \citealt{lsstSRD}), which is expected to begin survey operations in 2022, will discover $~10^4$ \SNeIa{} with high-quality light curves in its deep-drilling fields, as well as over a million \SNeIa{} with sparser light curves in the wide-fast-deep survey. 

Calibration of astronomical images is fundamentally the transformation of a number of ADU (Analog/Digital Units) from a source in a CCD image to a top-of-the-atmosphere brightness. This process has undergone many different iterations throughout the last 20 years of wide area astrophysical sky surveys. We briefly summarize these below.

The Sloan Digital Sky Survey (SDSS,\citealt{York}) made many innovations for the calibration procedures of wide-area sky surveys. They developed the $ugriz$ filter system \citep{SDSSSystem} that has been used with minor variations by many other surveys, including DES \citep{Flaugher15}, PS1 \citep{PS1System} and SNLS \citep{SNLSSystem}. The Ubercal method \citep{Ubercal} accounted for the flat field variation and amplifier gain variation while absorbing the atmospheric effects into a linear (in magnitude) airmass correction. This method made use of repeated observations of stars during the survey to achieve 1\% relative calibration (consistency in the natural magnitude system) across the survey footprint. Since Vega was too bright to be observed by SDSS, they tied their absolute photometry to the AB system \citep{AB}, a hypothetical flat reference spectrum which has a constant value of 3631 Jy (1 Jy = $10^{-26} \frac{W}{m^2 Hz}$) as would be measured at the top of the atmosphere. The AB system provides a more practical path to apply the absolute calibration through observations of fainter flux standards like BD+17-4708, which can be observed by large survey instruments without saturating the CCDs.

PS1 improved on the Ubercal method that SDSS used for its relative calibration by adopting a different survey strategy \citep{PS1Ladder}. This included larger areas of overlap between exposures and spacing repeat observations of a field both on 15 minute timescales within a night and at 6 month separations. These overlaps enabled PS1 to obtain high-quality calibration on nights with poor conditions as explained in \citet{PSCal1_5}. PS1 used their improved photometry and overlaps of their fields with those of SDSS to recalibrate SDSS to PS1-levels of precision using a method called Hypercalibration \citep{Hypercal}.

To further improve the absolute calibration, PS1 measured the full transmission function including instrument (telescope + CCD) and atmosphere. They measured the instrumental transmission using in-dome monochromator scans of the telescope and CCDs without filters, and they also utilized the vendor scans of the filter throughput \citep{PS1System}. For the atmospheric component, they included MODerate resolution atmospheric TRANsmission (MODTRAN, \citealt{MODTRAN}) models in their method to allow for specific contributions from aerosols, water vapor, and ozone to the linear airmass extinction. PS1 used repeated observations of many HST CalSpec \citep{CalSpec} standard stars inside of the footprint to tie its photometry to the AB system \citep{S15}. Hereafter ``transmission" refers to the full instrumental + atmospheric transmission function unless otherwise specified.

For the Dark Energy Survey \citep{DES} at the Cerro Tololo Interamerican Observatory (CTIO), a new calibration method has been developed based on a forward modeling approach. This method is called the Forward Global Calibration Method (FGCM, \citealt{B18}, B18 hereafter). While SDSS and PS1 account for the effect of the atmosphere averaged over each night, FGCM models the full DES transmission function for each CCD and each exposure, thus accounting for dependence of the transmission function on focal plane position as well as its time variation. FGCM uses approximately bimonthly measurements of the system throughput in each passband for each CCD (DECals, \citealt{DECals}).  More information about the variation in the instrumental transmission function across the focal plane is obtained from star flats as described in \citealt{DESY1Phot}. To monitor atmospheric changes, FGCM uses data from a GPS receiver at CTIO \citep{GPS, Flaugher15}. The data is analyzed by SuomiNet\footnote{http://www.suominet.ucar.edu}, which provides measurements of atmospheric precipitable water vapor (PWV) in 30 minute time windows. It uses this information in conjunction with data from bright stars in normal DES observations. FGCM achieves relative calibration at the $\sim 4.5$~mmag level. This is based on a comparison of the DES catalogs to those of Gaia DR2 averaged over small patches of sky and is likely an upper bound for the uncertainty.

FGCM determines its absolute calibration by comparing observed magnitudes of the CalSpec standard C26202  with the ``synthetic" magnitudes obtained by multiplying the CalSpec spectrum with each filter's model FGCM transmission function. C26202 is located within one of the DES-SN deep fields that has been observed over 100 times during the survey and it is faint enough to not saturate in most of the exposures. The flux scaling for each exposure, usually expressed as the logarithmic zeropoint, is obtained by integrating the product of the exposure's transmission function with a reference spectrum. Despite using C26202 to determine the absolute calibration, DES uses the flat AB spectrum as its reference.

This zeropoint is precisely correct only if the source SED is the reference spectrum or if it is observed under the exact conditions that define the reference transmission functions. The reference transmission functions are chosen during the FGCM fitting process to represent the average DES transmission function in each band. For other sources and other observing conditions, the optimal calibration requires additional corrections that depend on the SED of the source being observed. We call these ``chromatic corrections."

There are two observational effects that contribute to the need for chromatic corrections. First, the atmospheric transmission as a function of wavelength varies between observations. Those variations are illustrated in the uppermost panel of Fig.~3 from \citet{L16} (hereafter, L16). This figure shows the ratio between the transmission functions at PWV = 3 mm and PWV = 10 mm. This plot shows a maximum of 50\% fractional variation in the transmission function in $z$-band due to the PWV variation. Second, the Dark Energy Camera (DECam \citealt{Flaugher15})  filter transmission function varies across the focal plane as shown in Fig.~6 of L16. This figure shows a shift of the edge of the $i$-band transmission function of up to 6 nm as a function of distance from the center of the focal plane. 

\begin{figure}[ht!]
\centering
\includegraphics[width=8cm]{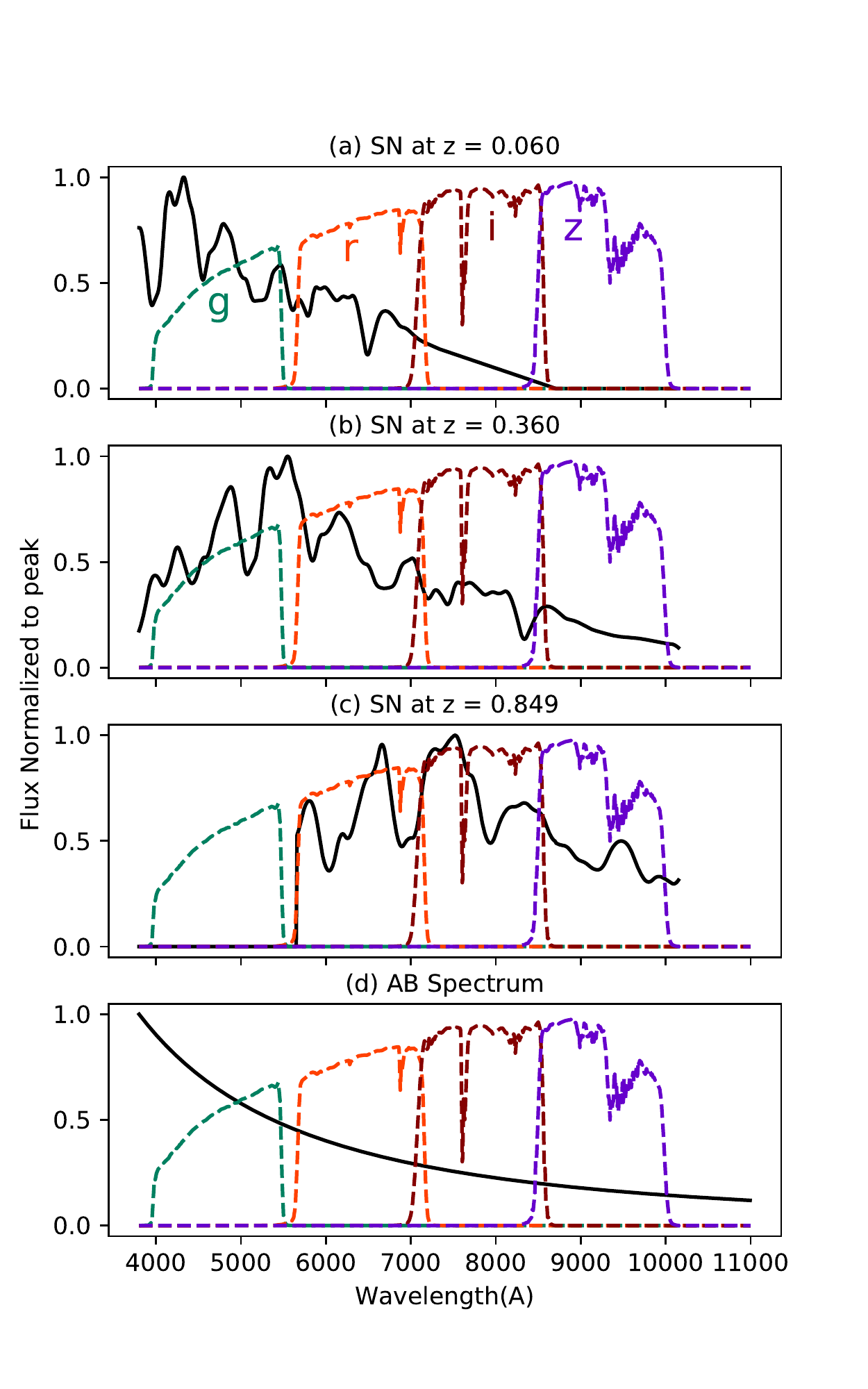}

\caption{\SNIa{} SEDs at peak brightness for (a) low, (b) intermediate, and (c) high
redshift model spectra from SALT2 \citealt{SALT2}, and (d) the AB spectrum, which is proportional to 1/$\lambda^2$. Overplotted on all spectra are the DES $griz$ standard bandpasses.}
\label{specEx}
\end{figure}

Color differences between astrophysical point sources and the reference standard affect the size of these chromatic corrections. This is particularly important for supernovae, as supernova SEDs are much redder than the reference standard, are very diverse, have strong broad features, vary with time, and vary significantly in color due to the wide range of redshifts observed as well reddening due to dust in the SN host galaxy. The variation of \SNIa{} spectra with redshift is shown here in Fig.~\ref{specEx}. The objective of this paper is to demonstrate the application of the chromatic corrections to DES-SN data and characterize the effects of the corrections in single-epoch photometry, light curves, and cosmology.

The outline of the paper is as follows. The formalism of chromatic corrections is described in \S \ref{CCMethod}. We describe the dataset to which these SED-based corrections are applied in \S \ref{sample}. We show the method with which the \SNIa{} light curves are fit, including the application of the chromatic corrections, in \S \ref{fitting}. In \S \ref{results}, we show our results including: demonstrating the effect of the corrections on the single-epoch photometry (\S \ref{seeffects}); the effect on the nuisance parameters $\alpha$ and $\beta$ as well as a cross-check of those parameters with those of Pantheon in \S \ref{nuisance}; the effect on the supernova light curves and cosmology (\S \ref{lccosmoeffects}); and the effect of the corrections on \SNIa{} simulated on individual CCDs (\S \ref{ccdeffects}). Finally we examine several cross checks on our analysis in \S \ref{crosschecks}.

\section{Methods + Data Sample}

\subsection{Application of Chromatic Corrections}
\label{CCMethod}
Here we describe the exact form of the chromatic corrections and the manner in which they are applied to the supernova photometry.

The typical definition of the magnitude, $m_{ b}$, in a band, $b$, of a source with flux (photon counts normalized by telescope aperture and exposure time), $F_{ b}$, in an image with zeropoint, $ZP_{ b}$, is

\begin{equation}
m_{b} = -2.5\log_{10}(F_{b}) + ZP_{b}.
\label{eqmag}
\end{equation}

In the AB system, this definition can be further expanded such that:

\begin{equation}
\begin{split}
m_{b} = -2.5\log_{10}\int_0^{\infty} \Fsrc(\lambda) \Phitot \lambda^{-1} d\lambda  \\
+ 2.5\log_{10}\int_0^{\infty} \Fref(\lambda) \Phitot \lambda^{-1} d\lambda,
\end{split}
\label{eqmag2}
\end{equation}

\noindent where $\Fsrc(\lambda)$ is the SED (in units of $\frac{\rm W}{\rm m^2 Hz}$) of the source object being observed, and $\Fref(\lambda)$ is the SED of the reference object for the photometric system. For DES, we use the AB spectrum (Fig.~\ref{specEx}d). $\Phitot$ is the dimensionless total transmission function.

This definition of the magnitude forms the basis for the chromatic corrections in L16:

\begin{flalign}
\delta_b^m =& m_{\rm std} - m_{\rm obs} =   \nonumber \\ 
-&2.5\log_{10}\frac{\int_0^{\infty} \Fsrc(\lambda)\Phiatmobs\Phiinstobs\lambda^{-1}d\lambda}{\int_0^{\infty} \Fsrc(\lambda)\Phiatmref\Phiinstref\lambda^{-1}d\lambda} \label{eqCC} \\ \nonumber
+& 2.5\log_{10}\frac{\int_0^{\infty} \Fref(\lambda)\Phiatmobs\Phiinstobs\lambda^{-1}d\lambda}{\int_0^{\infty} \Fref(\lambda)\Phiatmref\Phiinstref\lambda^{-1}d\lambda}.
\end{flalign}

In Eq.~\ref{eqCC}, $m_{\rm std}$ is the "standard" magnitude of the object being observe transformed as though it was observed under the reference conditions, $m_{\rm obs}$ is the magnitude that was observed under the actual conditions, $\Fsrc$ and $\Fref$ are the same as in Eq.~\ref{eqmag2}, $\Phiatmobs$ and $\Phiinstobs$ are the atmospheric and instrumental\footnote{telescope, instrument, filter, and CCD} components of $\Phitot$ at the location of the source on the focal plane from Eq.~\ref{eqmag2} such that $\Phiatmobs\Phiinstobs = \Phitot $~, and $\Phiatmref$ and $\Phiinstref$ are the reference atmospheric and DECam transmission functions within a given band, b, respectively. The reference transmission functions are chosen during the FGCM process to represent the most probable conditions over the course of the survey (see Fig.~4 in B18). 

Due to this choice of reference transmission, the average chromatic correction for a single object over an infinite number of observations should trend to zero. However, \SNeIa{} are time varying and the shape of the light curve is important for standardization. Therefore, trends in atmospheric parameters that depend on time (\textit{e.g}~seasonal variations, El Ni\~no, and degradation of the primary mirror) could produce effects that will not average to zero. The light curve sampling requirements result in non-uniform sampling of events over the course of the survey and therefore seasonal variations in atmospheric properties could potentially result in chromatic corrections whose effect on \SNIa{} distance does not average to zero.

The correction in Eq.~\ref{eqCC} is defined so that it is equal to zero for observations of the reference source with the reference transmission function. The atmospheric transmission functions are informed by our PWV measurements and the DECam transmission functions are measured by the DECal scans with additional input on the focal plane variation from star flats. The correction is added to the zeropoint based on the SED of the source.

These chromatic corrections are an improvement over the previous linear atmospheric correction\footnote{ http://classic.sdss.org/dr7/algorithms/jeg\_photometric\_eq\_dr1.html } in two major ways. First, they account for variation in the atmospheric conditions over the course of each night of observing whereas the linear correction coefficients were fit nightly. Second, the chromatic corrections incorporate SED information allowing for the correction of non-blackbody spectra and objects whose spectra have strong features

Using a small data sample, L16 shows that the effect of these chromatic corrections on \SNeIa{} can be as large as 10 mmag (1\%) in $z$-band and several mmag in $r$ and $i$ bands for high redshifts and large atmospheric water vapor.  This study illustrates that for SEDs that differ significantly from the reference, the chromatic corrections can be comparable or larger than to the non-uniformity of the calibration ($\approx$ 4.5 mmag). 

The middle row of panels in Fig.~7 from L16 shows that the variation in the corrections described in Eq.~\ref{eqCC} matches the observed variation in stellar magnitude vs. stellar color to mmag precision. The typical color range observed in \SNeIa{} is $0.5 < g-i < 3.5$, which includes the entire range of that figure. This test demonstrates that chromatic corrections improve the calibration for sources whose SED differs from the reference SED.

\subsection{Data Sample}
\label{sample}
The Dark Energy Survey includes a 5000 $\rm deg^2$ (``wide") survey \citep{DESY5} and a 27 $\rm deg^2$, time domain, supernova survey \citep{B12,Diffimg}, which are run concurrently between August and February beginning in 2013 and ending in 2018. The wide survey alone will continue operations into 2019. The wide survey is conducted in 5 bands ($grizY$) of which the 4 bluest bands ($griz$) are used in the supernova survey. Survey observations are conducted on the Victor Blanco 4m telescope using the Dark Energy Camera (DECam) at the CTIO in Chile. 

\begin{figure*}[]
\centering
\includegraphics[width=18cm]{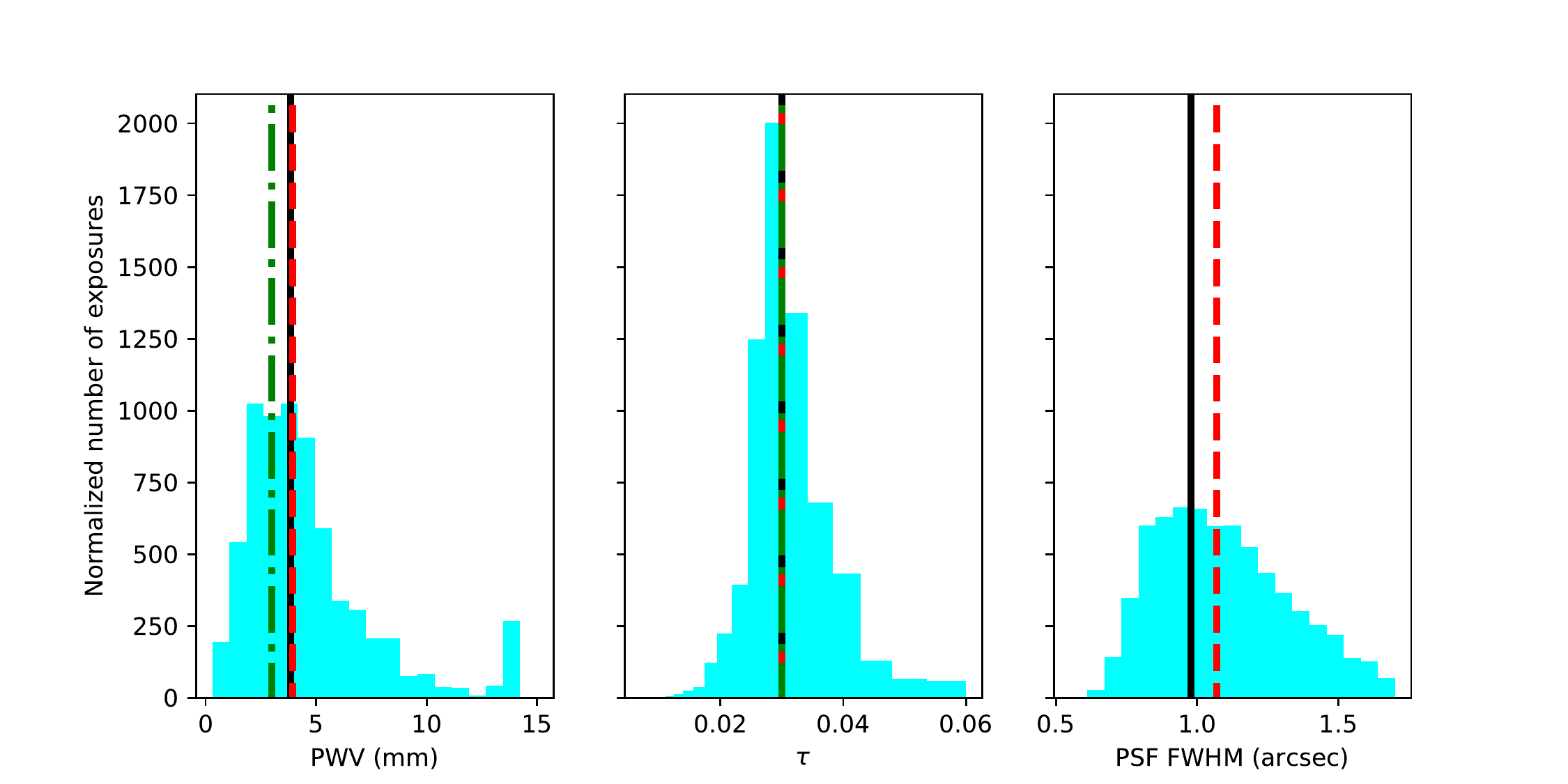}
\caption{The distribution of atmospheric precipitable water vapor (PWV), optical depth due to aerosols ($\tau$), and PSF FWHM for DES observations of the supernova fields during the first three seasons. The black solid and red dashed lines represent the median conditions over the first three seasons of the survey for DES and DES-SN respectively, and the green dotted and dashed line represents the reference atmosphere (there is no standard PSF size). All four bands ($griz$) are included.}
\label{atmDist}
\end{figure*}

The supernova fields are observed when the predicted point spread function (PSF) is above 1.1$\arcsec$ or when a field hits a ``deadman" trigger meaning that it has not been observed for 7 days.  The atmospheric conditions of the supernova survey are illustrated in Fig.~\ref{atmDist}, whose three panels shows the distribution of PWV, atmospheric optical depth  due to aerosols ($\tau$), and PSF respectively. While PWV and $\tau$ are comparable to the median DES wide area conditions, the median PSF is about a tenth of an arcsecond above the median PSF of the wide area survey.

Atmospheric parameters PWV and $\tau$ were computed by B18 for exposures satisfying quality requirements for the wide-area, and thus 10\% of supernova survey observations do not have the atmospheric parts of the correction. However, all exposures are corrected for instrumental transmission variation. Atmospheric information for all exposures will be included in a future paper that will cover the calibration of the full five seasons of DES.

The DES SNe are discovered in the ``real-time" difference imaging pipeline (DIFFIMG, \citealt{Diffimg}), where deep coadded template images are subtracted from each supernova survey image. In this paper we use \samplesize{}  spectroscopically confirmed \SNeIa{} discovered from the first three years of DES-SN. The spectroscopic selection of the sample is described in D’'Andrea et al (2018, in prep.). The 2\% calibration uncertainty for DIFFIMG photometry is sufficient for SN discovery and monitoring, but is not sufficient for the cosmology analysis. Therefore, DES has developed a version of ``scene modeling" photometry (SMP, \citealt{Brout18}), originally developed by SDSS  \citep{SMP} and later used by SNLS \citep{SMP2}, for use in the offline analysis with the goal of achieving sub-percent precision. In this analysis we are using the SMP photometry.

To improve the cosmological parameter determination, the spectrscopically confirmed subset of the  DES-SN sample is combined with \NLowz{} low redshift (\lowz{}) supernovae from surveys including CFA3, CFA4, and CSP. This sample is taken from the Pantheon analysis \citep{Pantheon} with additional cuts described in \citealt{KEYPAPER}. We do not apply chromatic corrections to the \lowz{} sample because we do not have the information necessary to make these corrections. Instead, we use the original survey calibration. The combination of the DES-SN sample and the \lowz{} sample is referred to as ``\samplename{}."

In order to study the effect of chromatic corrections with large statistics in all areas of parameter space (\textit{e.g.}~SN parameters like redshift, color, and stretch as well as observing conditions like $\tau$ and PWV), we utilize a \samplename{}-like sample produced by the simulation code in the SuperNova ANAlysis (SNANA\footnote{snana.uchicago.edu for manual and other information}, \citealt{SNANA}, \citealt{Kessler18}) software package. We use this simulation to generate a large number of SEDs, approximately 120x the size of the \samplename{} sample, for which we can assess the impact of chromatic corrections. 

These simulated supernovae are generated using the color and stretch distributions of \citet{SK16}, the volumetric rate from \citet{P12}, the spectroscopic selection function from \citealt{DAndrea18} in Prep., host galaxy library from \citet{G16}, and the intrinsic scatter model from \citet{G10,K13}. This simulation uses randomly chosen sky coordinates over the supernova fields, selects a random CCD from the focal plane, and uses DES observation dates. The date and focal plane location are used to determine chromatic corrections (Eq.~\ref{eqCC}) in the same manner as for the data. 

\begin{figure*}[]
\centering
\includegraphics[width=18cm]{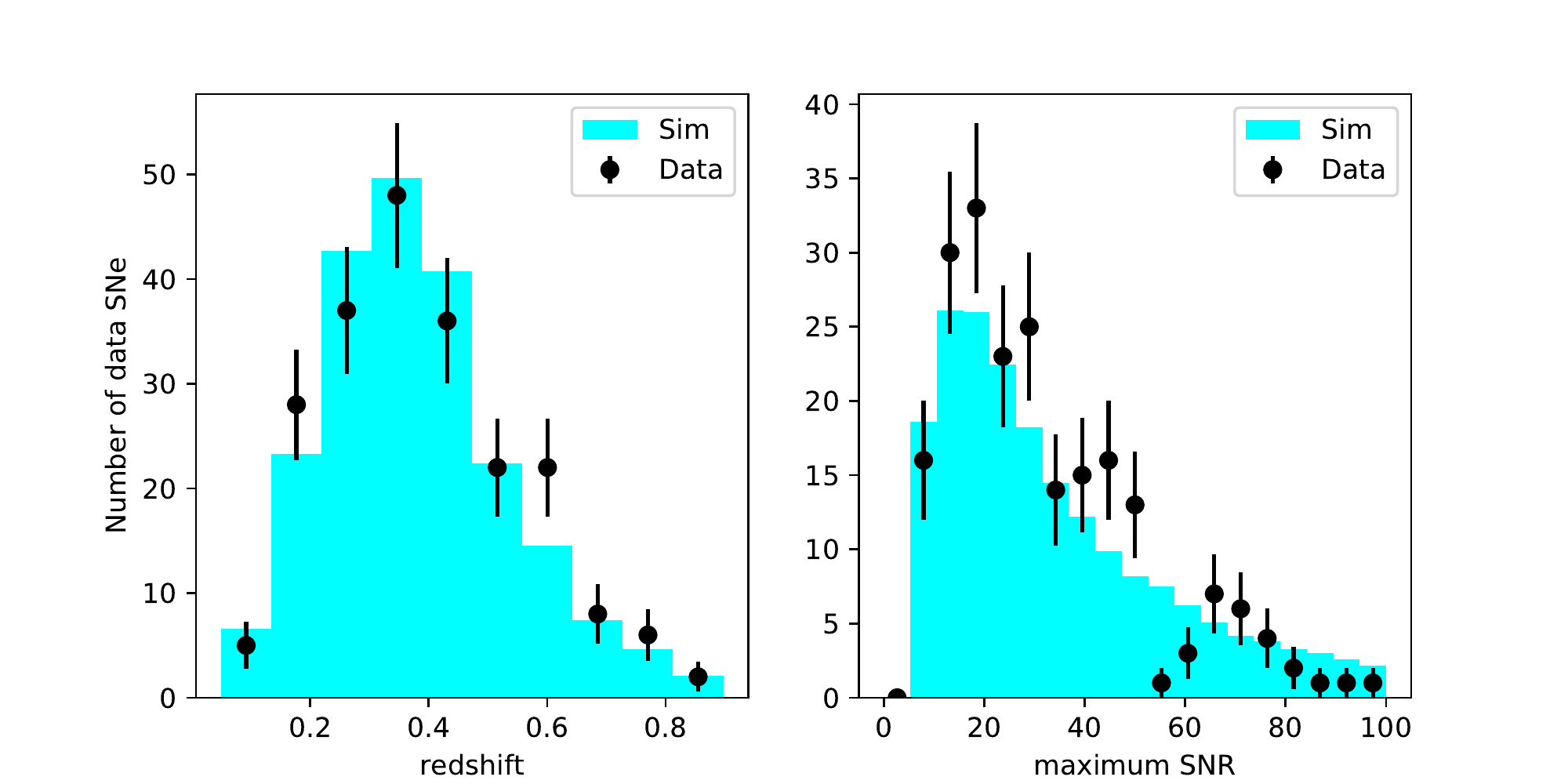}

\caption{The distribution of redshift (left) and maximum signal to noise (right) for \SNeIa{} in the DES-SN data sample  (black circles with error bars) and simulation (light blue bars).}
\label{simDataComp}
\end{figure*}

Fig.~\ref{simDataComp} shows the redshift and maximum signal to noise ratio (SNR) distributions for the DES-SN sample. The simulations agree well with the data for the DES-SN sample. A similar plot for the \lowz{} sample is shown in Fig.~7. of \citep{Kessler18}

\subsection{Light-Curve and Cosmology fitting}
\label{fitting}
The SNANA software package provides light-curve fitting code using the Spectral Adaptive Lightcurve Template 2 (SALT2) model first developed by \citet{SALT2}. We use the most recently trained SALT2 model that was developed for the Joint Light-curve Analysis (JLA, \citealt{JLA}). However, to see the effect of the chromatic corrections on $z$-band in the lowest redshift DES supernovae, we use the Near Infrared (NIR) extension of this model from \citet{H17}. The light-curve fitting code determines the stretch ($x_1$), color ($c$), amplitude ($x_0$), and time of peak brightness ($t_0$) for each supernova light-curve, both for the \samplename{} data sample and the simulated sample described in \S \ref{sample}. Distance moduli are calculated by the Tripp estimator \citep{Tripp}:

\begin{equation}
\mu = M_0 + m_B + \alpha x_1 - \beta c, 
\end{equation}

\noindent 
where $m_B = -2.5\log_{10}(x_0)$, $\alpha$ is the stretch-magnitude standardization parameter, and $\beta$ is the color-magnitude standardization parameter.

In the first step of the analysis, the light curves are fit without chromatic corrections to determine the SED at each epoch using the SALT2 spectral model. Next, the light curves are fit with corrections (Eq.~\ref{eqCC}) applied.

After light curve fitting, the standardization parameters $\alpha$ and $\beta$ and a  Hubble diagram which is corrected for biases due to selection effects and light curve fitting are determined simultaneously from a global fit to the set of \samplename{} light curve parameters ($c$, $x_1$, $m_B$). This global fit is performed with the BEAMS with Bias Correction (BBC, \citealt{KS17}) formalism with 20 logarithmically spaced redshift bins from 0.01 to 0.85.

The binned distances and uncertainties are passed to wFit, a fast $\chi^2$ minimization program using MINUIT \citep{MINUIT}, which outputs marginalized cosmology parameters $w$ and $\Omega_m$ based on a $w$CDM model, a flat universe with varying dark energy equation of state parameter, $w$, and cold dark matter. These parameters are obtained with priors from BAO \citep{EisensteinBAO} and CMB \citep{WMAP}. The cosmological parameters are blinded so that we only examine differences due to the chromatic corrections. These simplifications are used because they are significantly faster and sufficiently accurate for differential studies. However, we do not use these simplifications in the nominal \samplename{} cosmology analysis \citep{KEYPAPER}.

For each SN, we calculate the change in the BBC distance modulus $\mu$ and the three parameters $x_1$, $c$, and $m_B$ due to the chromatic corrections. The light-curve fit parameters $x_1$ and $c$ are multiplied by the nuisance parameters $\alpha$ and $\beta$ to give them the same units (mag) as $\mu$ and $m_B$. $\alpha$ and $\beta$ are fit separately with BBC before and after corrections are applied; however, they do not significantly change due to the corrections. Therefore, we adopt a single value for alpha and beta when calculating the differences. These differences are defined below:

\begin{equation}
\dmu = \mu_{\rm noCorr} - \mu_{ \rm corr}
\label{eqdmu}
\end{equation}

\begin{equation}
\dax = \alpha x_{1,\rm noCorr} - \alpha x_{1, \rm corr}
\label{eqdax}
\end{equation}

\begin{equation}
\dbc = \beta c_{\rm noCorr} - \beta c_{ \rm corr}
\label{eqdbc}
\end{equation}

\begin{equation}
\dmB = m_{B, \rm noCorr} - m_{B, \rm corr}.
\label{eqdmB}
\end{equation}

We characterize the $\Delta$ parameter dependence on redshift using a linear fit ($\Delta$ vs.~redshift) to the unbinned SN sample in order to obtain a simple one parameter quantification of the effect of the corrections. These values and their slopes can be seen in Figs. \ref{dataPanel}, \ref{simPanel}, \ref{SlopeIdeogram}, and \ref{CCDPanel}. Since the fitting uncertainty on the slopes of the best fit lines does not account for correlations (\textit{e.g.}~between $x_{1,\rm noCorr}$ and $ x_{1,\rm corr}$), the uncertainty is determined empirically. We generate 50 data-sized simulations of the \samplename{} sample. Then, after running those samples through the same analysis as the data, we collect the fitted values of the $\Delta$ parameter slopes vs. redshift and $\Delta$ cosmological parameters. We use the standard deviation of the slopes and cosmological parameters among the 50 simulations to estimate the uncertainty. We believe this is valid since the distribution of these 50 slopes is consistent with a normal distribution. We also use those uncertainty estimates for the larger simulated sample. However, for that larger sample we scale down the uncertainty by the square root of the ratio of the size of the larger sample to the size of the \samplename{} sample. Applying corrections based on these linear relationships is not a substitute for applying the full integrated correction to each supernova epoch. However, calculating the slopes is useful to check whether the simulated SNe change similarly to the data events as well as to check whether the redshift trend (or lack thereof) in parameter changes indicates that there should or should not be a cosmological parameter bias.

We define changes in the wFit output cosmological parameters $w$ and $\Omega_m$:

\begin{equation}
\dw = w_{\rm noCorr} - w_{ \rm corr}
\label{eqdw}
\end{equation}

\begin{equation}
\dOm = \Omega_{\rm m,  noCorr} - \Omega_{\rm m,  corr}.
\label{eqdOm}
\end{equation}

Following the method for determining the uncertainties on the slopes above, the uncertainties on $\dw$ and $\dOm$  are the standard deviation in these quantities from 50 \samplename{} sized simulations. These uncertainties are also scaled by the square root of the sample size.

\section{RESULTS}
\label{results}
We begin this section with the impact of the corrections on the single-epoch photometry and show the dependence on SN color, redshift, and atmospheric PWV. Next we show a comparison of the SALT2 nuisance parameters $\alpha$ and $\beta$ between this analysis and the Pantheon analysis, as well as the change in those parameters due to the chromatic corrections.  Finally, we present the changes in light-curve fit parameters, distance moduli, and cosmological parameters due to these corrections.

\subsection{Impact on single-epoch photometry}
\label{seeffects}

We apply the corrections described in \S \ref{CCMethod} to the \samplename{} sample and examine the effects on single-epoch photometry. Since the $z$ band includes water absorption lines, we present those results here. The $g$, $r$, and $i$ bands show a median chromatic correction consistent with zero at all values of redshift, PWV, and observed $r-i$ (for $g$ and $r$ band) or $i-z$ (for $i$ band) color. The standard deviation of all chromatic corrections in $g$, $r$, and $i$ bands respectively are 11.1, 3.3, and 4.4 mmag. 

\begin{figure*}[]
\centering
\includegraphics[width=18cm]{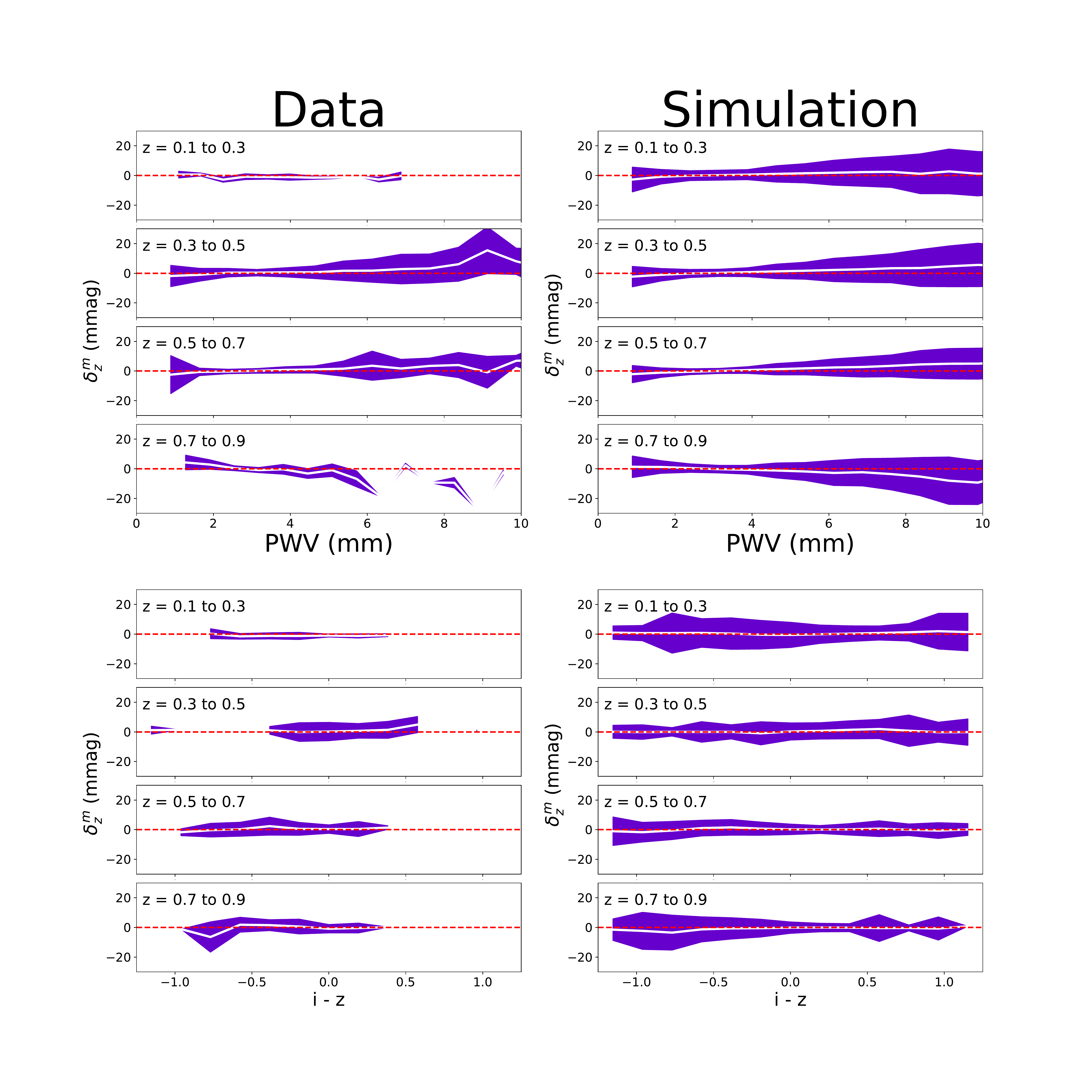}

\caption{For DES $z$ band, $\delta_z^m$ dependence on PWV (top) and $i-z$ color (bottom). Each set of 4 panels shows a different redshift range for data (left) and simulations (right). The white solid lines connect the median chromatic correction in each PWV/color bin, the red dashed line is zero, and the colored band represents the standard deviations within each bin.}
\label{CC-z}
\end{figure*}

 The upper left panel of Figure~\ref{CC-z} shows the average ($z$-band) chromatic correction as a function of PWV and $i - z$ color when applied to the \samplename{} sample. PWV and $i -z$ are divided into 10 evenly spaced bins over the range of observed parameter space. Those panels are further subdivided into panels based on the SN redshift. These plots include all SN epochs regardless of phase relative to peak brightness in the  model B band.

There is a trend of about $1$~mmag per mm of PWV at low redshifts and that trend reverses to $-1$~mmag per mm of PWV at highest redshifts. To see the $\delta_z^m$ effect with higher statistics, the upper right panel of Figure \ref{CC-z} shows a prediction using a simulation of 120 \samplename{} samples. This simulation confirms the trend observed in the data. There is no statistically significant trend with light-curve fit color in data or simulation. The data sample appears to have very low scatter in some PWV/color+redshift bins because it only has one or two events that fall in that bin. The simulated scatter is more representative of the true scatter in the chromatic corrections.

\begin{figure*}[]
\centering
\includegraphics[width=18cm]{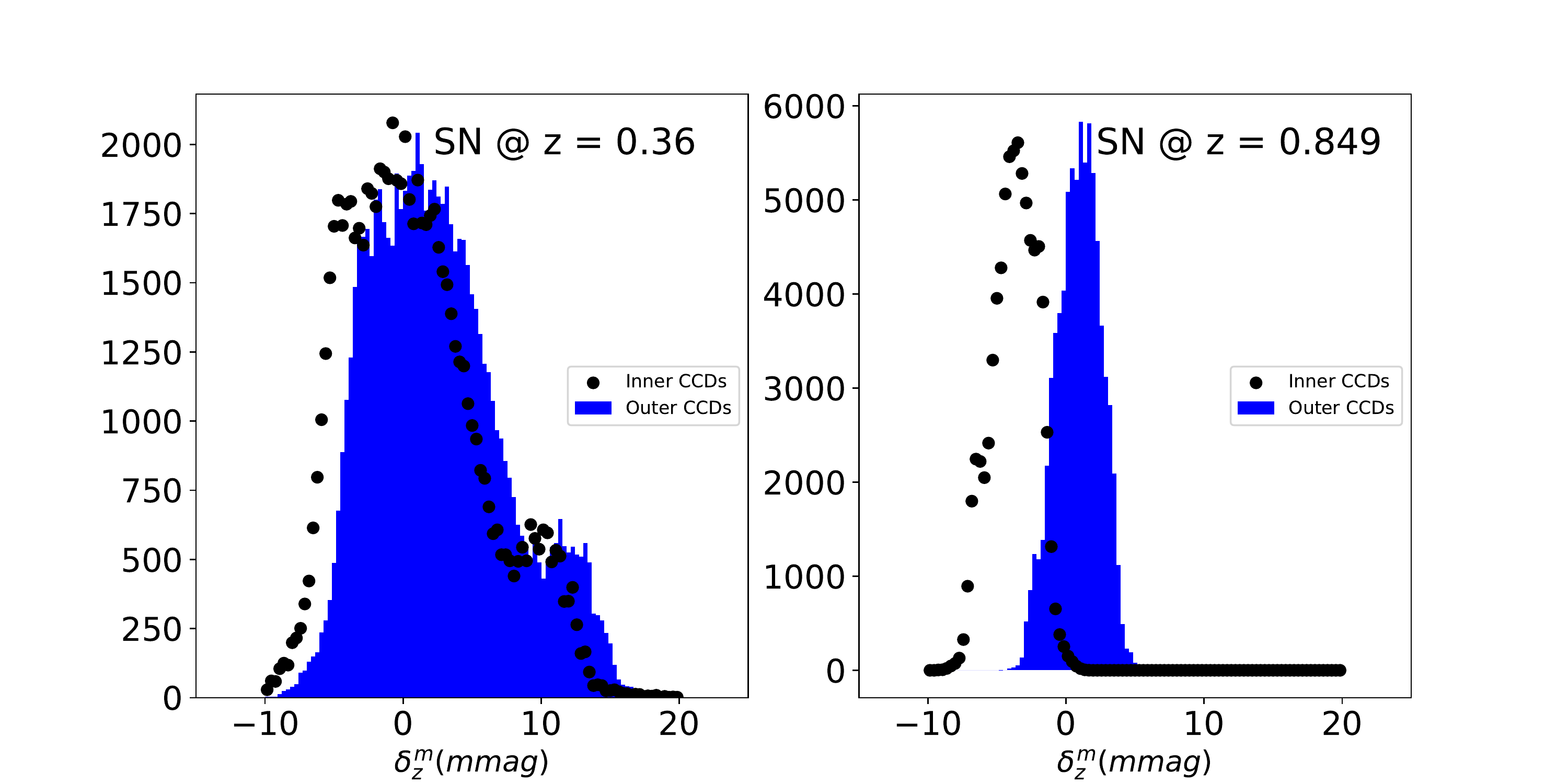}

\caption{For DES $z$ band, $\delta_z^m$ distribution for the SN spectrum in Fig.~\ref{specEx}~b (left) and Fig.~\ref{specEx}~c (right). Each set of panels shows a different set of CCDs (inner CCDs as black dots and outer CCDs as blue bands).}
\label{CorrDistroSample}
\end{figure*}

In order to further illustrate the effect of the chromatic corrections due to atmospheric and CCD variations, in Fig.~\ref{CorrDistroSample} we present the distribution of corrections for two selected SN SEDs integrated for each atmospheric transmission function observed during DES. These SN SEDs are from Figs.~\ref{specEx}~b and \ref{specEx}~c, with redshifts 0.36 and 0.85, respectively. We present two panels for each of the two sample SEDs: the first set of panels takes its instrumental transmission function from 6 interior CCDs and the other takes its instrumental transmission function from the 6 outer CCDs. The median of the chromatic correction distribution is significantly different when considering the inner CCDs vs. the outer CCDs and the shape of the distribution is much wider for the lower redshift SN than the higher redshift SN.

The width of the low-redshift chromatic correction distribution is driven primarily by PWV variations between 0.5 and 15 mm. Within the z-band wavelength range, the AB spectrum is nearly flat, while the SN spectra are significantly more tilted, and thus PWV variations, which affect the region near $\lambda \sim 9500 \angstrom$, have a larger effect on the AB spectrum.  The distribution for high redshift (lower panels) is much narrower due to the fortuitous bump in the exact location of the PWV feature. We have checked that without this bump, the width of the chromatic correction distribution is much larger and also reproduces the secondary peak that we observe in the low redshift distribution.

\subsection{Result for BBC fitted SALT2 nuisance parameters}
\label{nuisance}

Table \ref{tab1} shows the nuisance parameters $\alpha$ and $\beta$ from the BBC fits of the \samplename{} sample, as well as those from the Pantheon Sample. We compare these parameters to check our fitting method without unblinding the cosmological parameter fit. $\alpha$ and $\beta$ are statistically consistent between \samplename{} and Pantheon, and the chromatic corrections result in negligible shifts ($<1\%$). The table also includes the intrinsic scatter of supernova brightness ($\sigma_{\rm int}$) which is calculated as the amount of additional error that needs to be added during the fit to get the reduced $\chi^2$ to be equal to one. This value is also comparable to the scatter in Pantheon \citep{Pantheon}.

\begin{table}
\caption{BBC nuisance parameters for \samplename{} and Pantheon samples.} \label{tab1}
\begin{center}
\begin{tabular}{ |c|c|c|c| }
\hline
 Dataset & $\alpha$ & $\beta$ & $\sigma_{\rm int}$ \\
\hline
 DES Uncorrected& 0.144$\pm$ 0.008 &3.12 $\pm$ 0.104 &0.097 \\
\hline   
 DES Corrected\footnote{Chromatic corrections described in \S 2.1.}& 0.145$\pm$ 0.008 &3.11 $\pm$ 0.10 &0.097 \\
\hline
 Pantheon & 0.156 $\pm$ 0.006 & 3.02 $\pm$ 0.06 & 0.09 \\
\hline

\end{tabular} \par \bigskip

\end{center}
\end{table}

\subsection{Effect of chromatic corrections on light-curve fit parameters distances, and cosmology}
\label{lccosmoeffects}

Here we propagate both the \samplename{} and simulated samples through the analysis and show the effects of the chromatic corrections as a function of redshift on the fit parameters.

\begin{figure*}[]
\centering
\includegraphics[width=18cm]{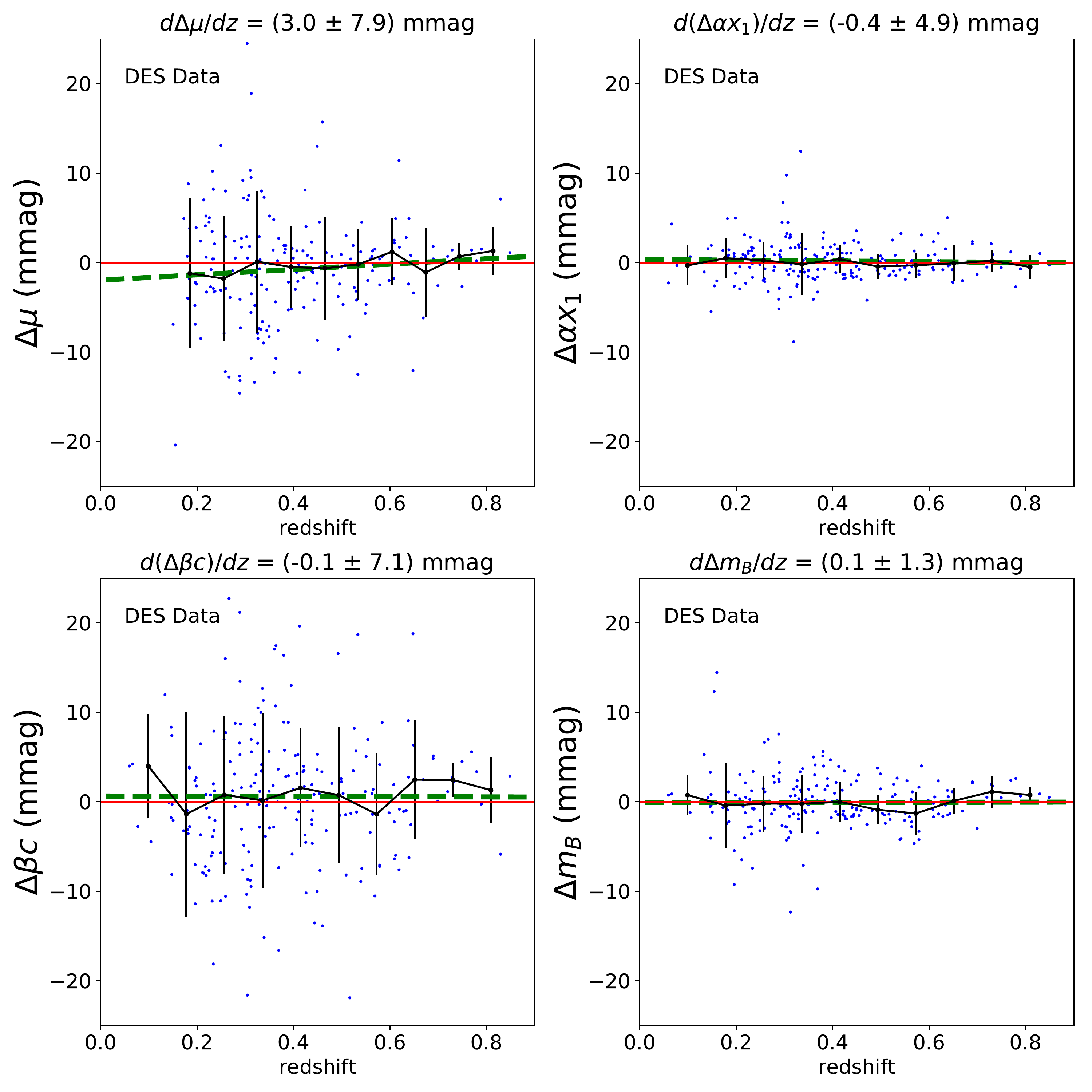}
\caption{The redshift dependence of $\dmu$, $\dax$, $\dbc$, and  $\dmB$ (Eqs.~\ref{eqdmu}, \ref{eqdax}, \ref{eqdbc}, and \ref{eqdmB} respectively) for the \samplename{} \SNIa{} sample. Each dot is an individual SN from the \samplename{} sample. The black line connects the median $\Delta$ for the SNe in each redshift bin, its error bars represent the standard deviation of the chromatic correction in each bin, the red solid line is zero, and the green dotted line is the best fit line whose slope and uncertainty are given above each panel.}
\label{dataPanel}
\end{figure*}

Figure \ref{dataPanel} shows the redshift dependence of the effect of the chromatic corrections on the measured distance modulus ($\Delta \mu$), as well as on the light-curve fit parameters ($\Delta x_1$,  $\Delta c$, and $\Delta m_B$) for the data. The slopes of the shift vs.~redshift given on the top of each panel show that each slope is consistent with zero.

\begin{figure*}[]
\centering
\includegraphics[width=18cm]{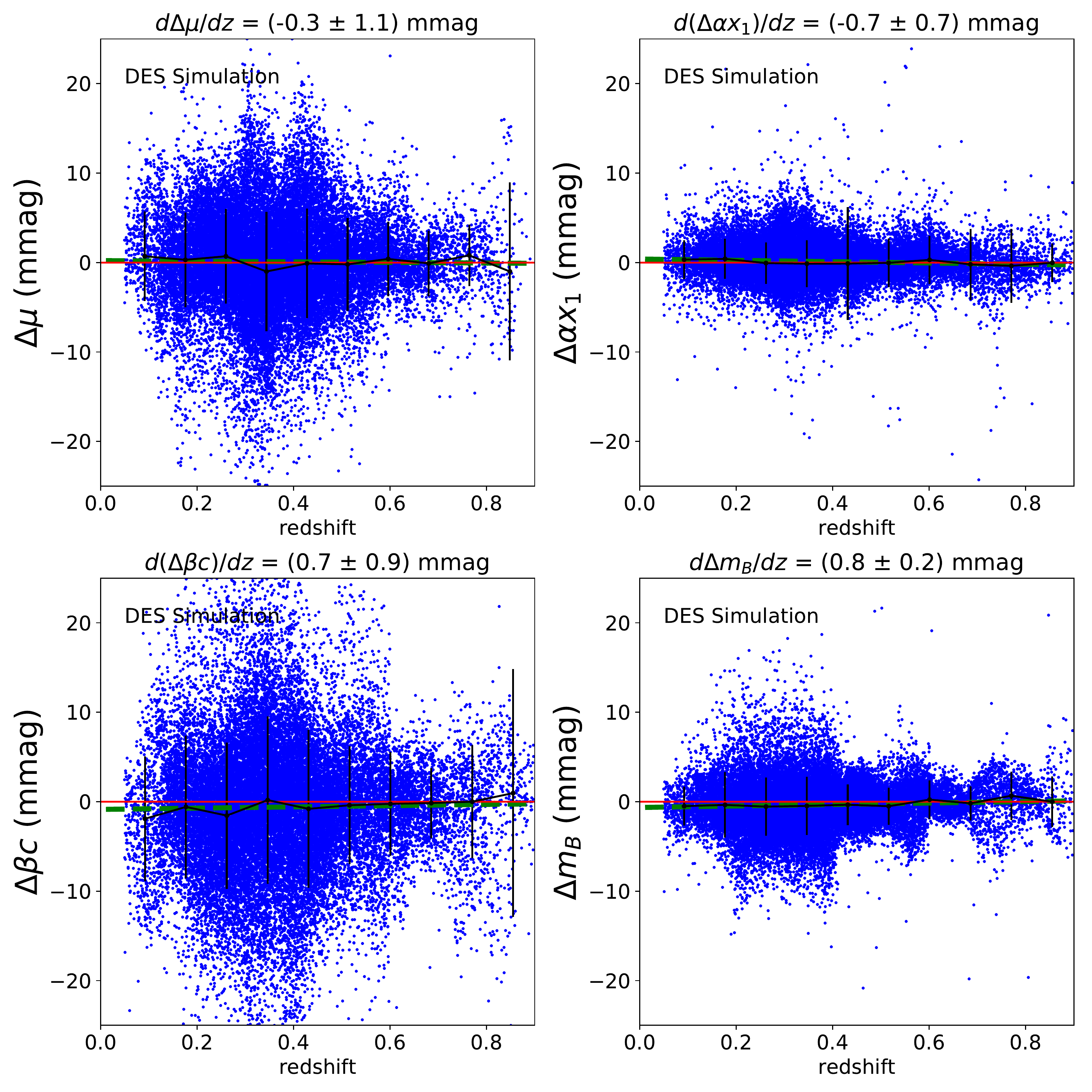}
\caption{Same as Fig.~\ref{dataPanel}, but for the simulated \samplename-like sample.}
\label{simPanel}
\end{figure*}

\begin{figure*}[]
\centering
\includegraphics[width=18cm]{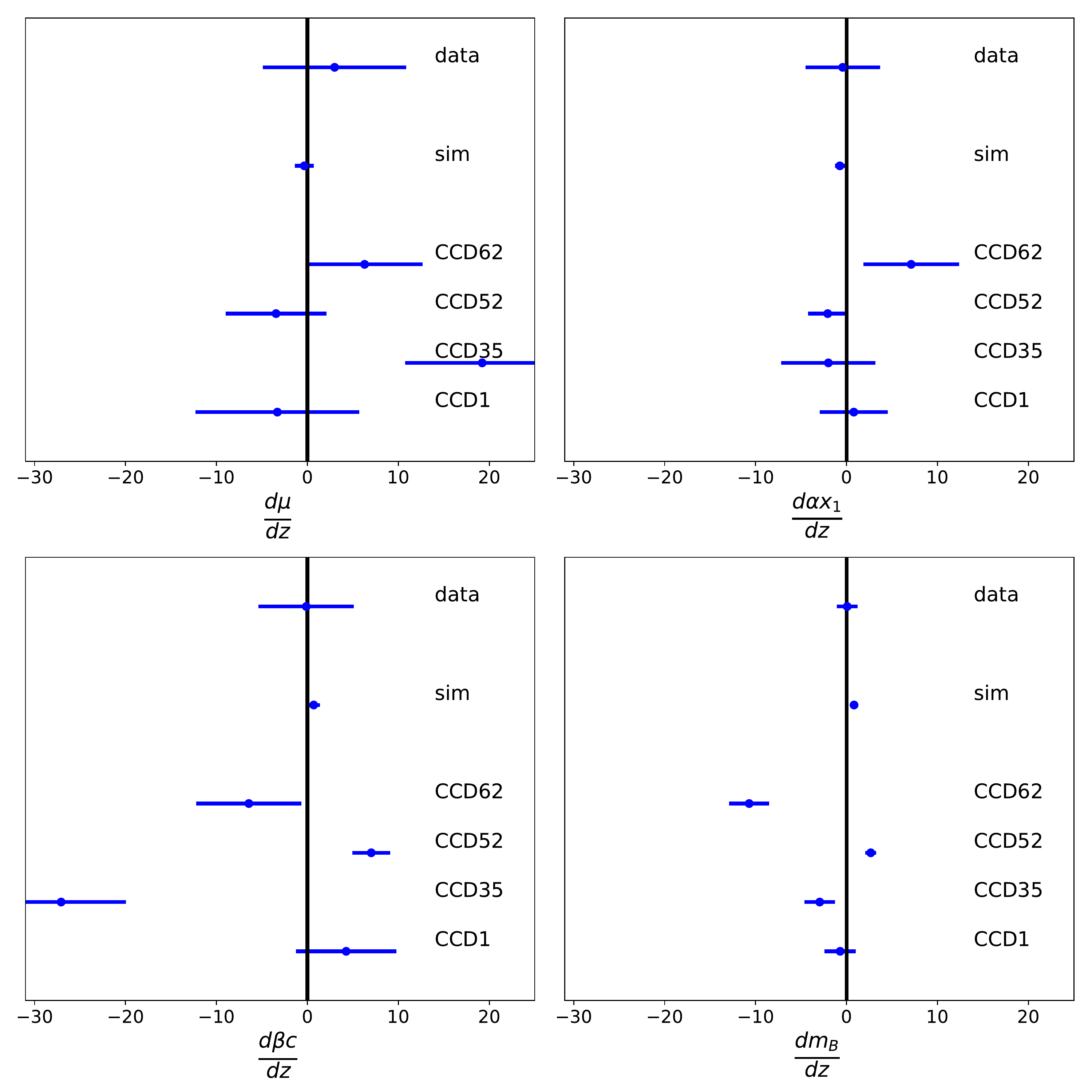}
\caption{The slope of the  $\dmu$, $\dax$ (upper right), $\dbc$ (lower left), and $\dmB$ (lower right) (Equations \ref{eqdmu}, \ref{eqdax}, \ref{eqdbc}, and \ref{eqdmB}) vs.\ redshift for the data sample, simulated sample, and subsets of the simulated sample for 4 individual CCDs.}
\label{SlopeIdeogram}
\end{figure*}

Figure \ref{simPanel} shows the same quantities as in Fig.~\ref{dataPanel}, but for the simulated \samplename{} sample, which has slope uncertainties that are almost
an order of magnitude smaller than those of the data. For the simulated SNe, $\dmB$ shows a nonzero slope with  3-$\sigma$ significance (0.6 $\pm$ 0.2 mmag). This effect does not propagate to any significant redshift trend in distance modulus vs. redshift. The remaining light-curve fit parameters have slopes vs.\ redshift that are consistent with zero at a 1-$\sigma$ level.  All of the slopes for the simulated sample parameters are consistent with the slopes of the data sample parameters as shown in the top two rows of each panel of Fig.~\ref{SlopeIdeogram}. While the mean of the chromatic corrections is small, the scatters in these plots exhibit the range of the chromatic corrections on the individually measured distance moduli and fitted light-curve parameters.

To examine the relative effects of the atmospheric and instrumental corrections, we made the corrections for the simulated sample using the standard atmosphere, zeroing out the atmospheric correction, and then we made a second set of corrections that are the differences between the full (atmospheric + instrumental) corrections and the instrumental only corrections. The effect of these corrections on distance modulus are shown in Fig.~\ref{AtmOnly} and Fig.~\ref{CCDOnly} below. 

\begin{figure*}[]
\centering
\includegraphics[width=14cm]{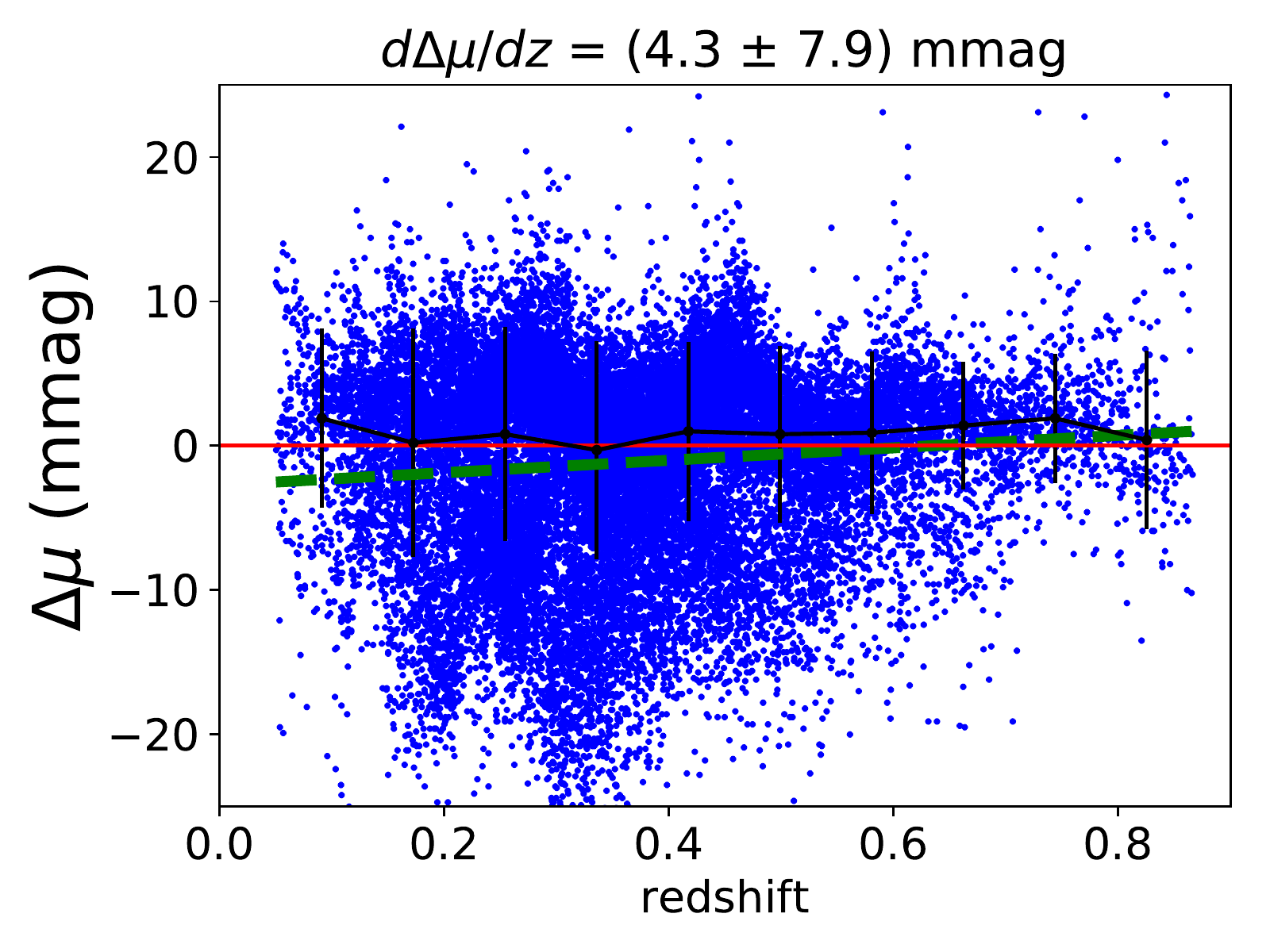}
\caption{Same as Fig.~\ref{simPanel} upper left panel, but for atmospheric corrections only, and only for the four chosen CCDs (1,35,52,62).}
\label{AtmOnly}
\end{figure*} 

\begin{figure*}[]
\centering
\includegraphics[width=14cm]{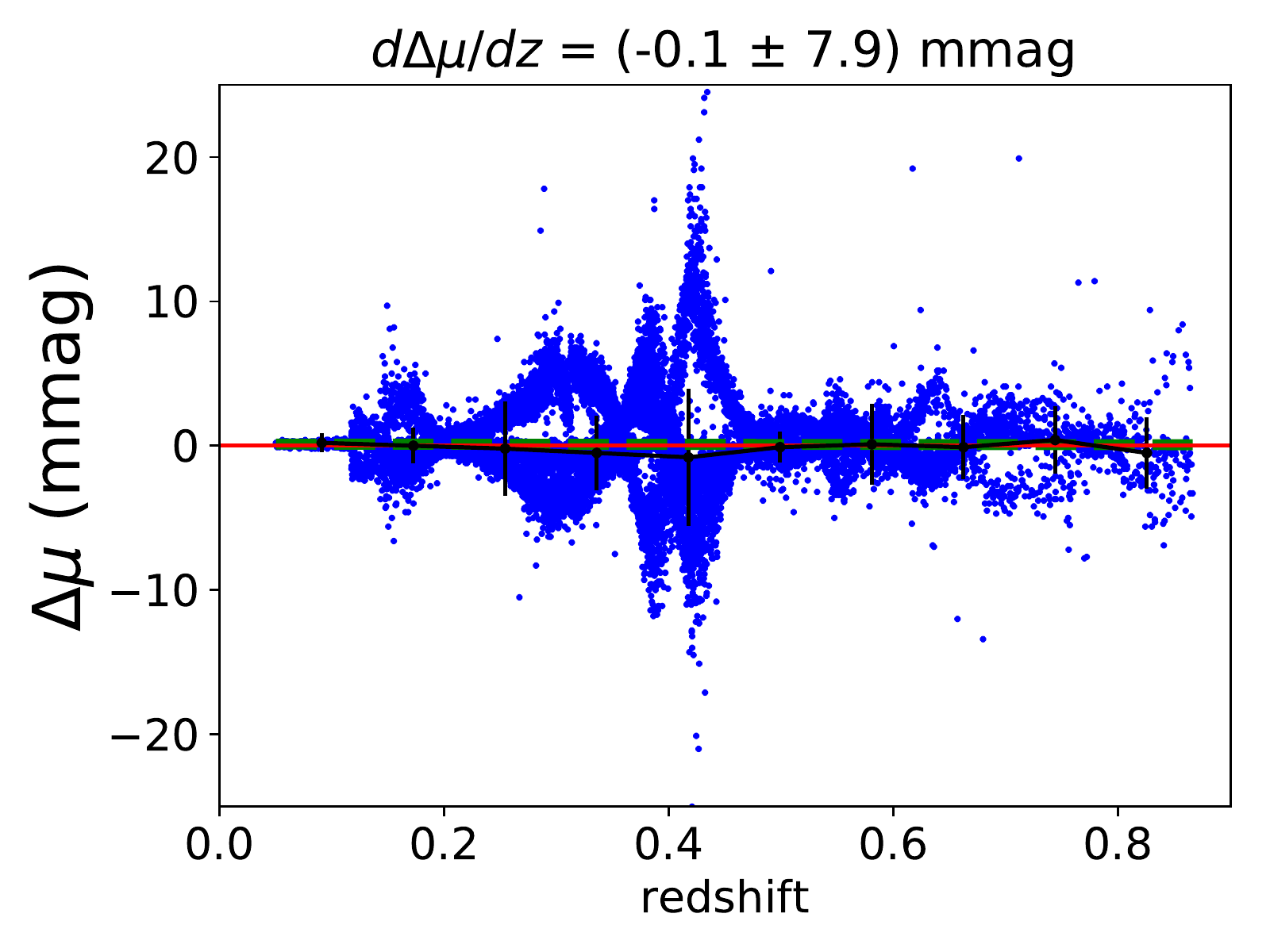}
\caption{Same as Fig.~\ref{AtmOnly}, but for instrumental corrections only.}
\label{CCDOnly}
\end{figure*} 

These figures (Fig.~\ref{AtmOnly} and \ref{CCDOnly}) show that the trend in distance correction vs. redshift is mostly due to the atmospheric effects, but the oscillatory features are mostly due to the instrumental effects. We have examined the trend of $\dmu$ vs. redshift for individual CCDs, and we find that the oscillations are present in each CCD and are not an artifact of stacking all of the CCDs.

\begin{figure*}[]
\centering
\includegraphics[width=18cm]{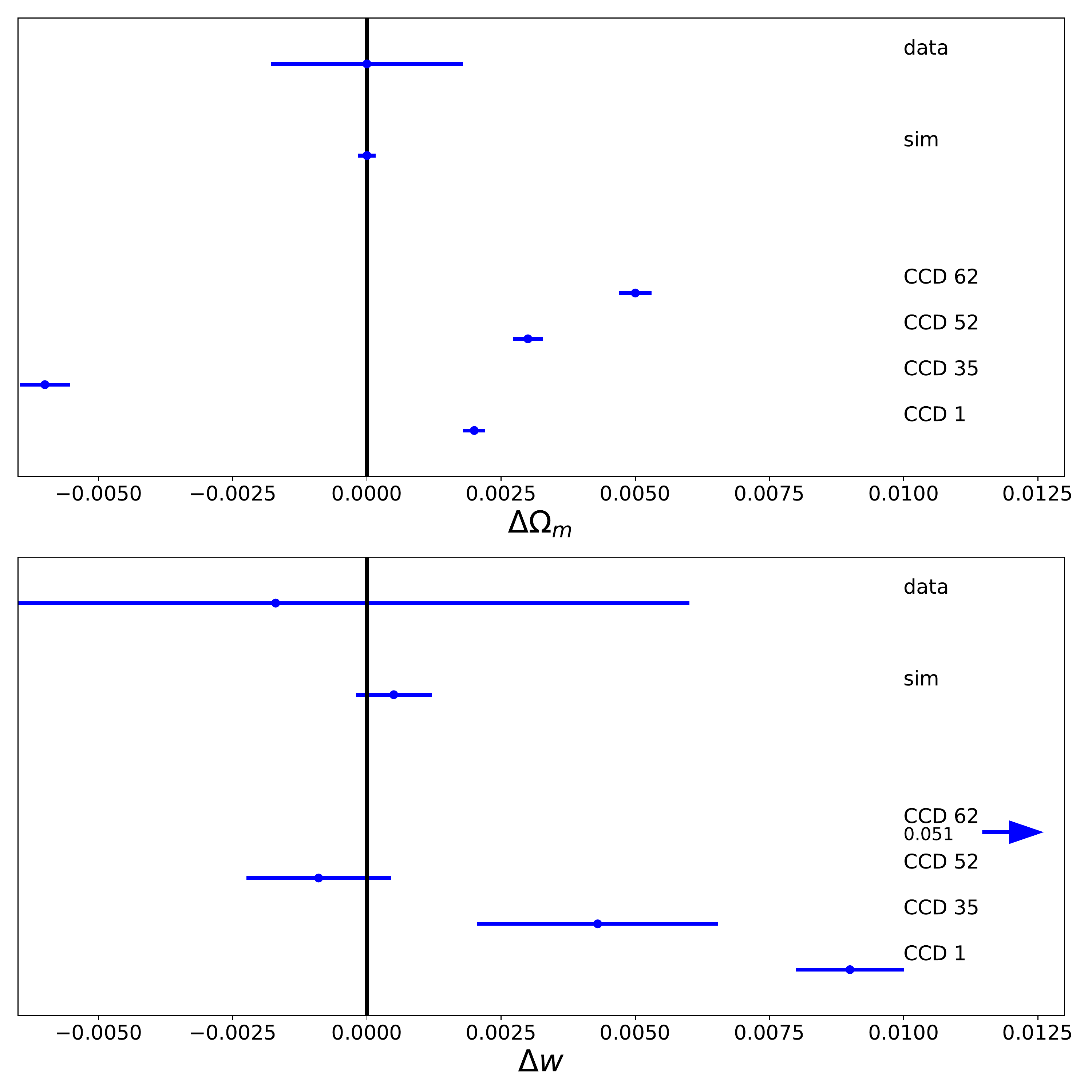}
\caption{$\dOm$ and $\dw$ (Equations \ref{eqdw} and \ref{eqdOm}) for the labeled samples. $\dw$ for CCD62 is out of view so its value is printed next to an arrow.}
\label{CosmoIdeogram}
\end{figure*}

For the data, the shifts in the cosmological parameters $\Omega_m$ and $w$ due to the chromatic corrections ($\dOm$ and $\dw$ as in Equations \ref{eqdw} and  \ref{eqdOm}) are $\dw = \wcorr$ and $\dOm = \omcorr $. These changes are consistent with our simulated results where the mean change in $w$ over our 50 simulated \samplename{}-sized simulations is $\fiftysimwcorrmean$ with an standard error in the mean of $\fiftysimwcorrerr$. Similarly, the mean $\Omega_m$ change is $\fiftysimomcorrmean$ with a standard error in the mean of $\fiftysimomcorrerr$. The simulation results are consistent with zero as expected. They are also consistent with the data based on our limited sample size. The results are visualized in the top two rows of each panel of Fig.~\ref{CosmoIdeogram}.

For an \SNIa{}-only analysis, we find $\dOm$ and $\dw$ are 0.005 and -0.0294. However, since the shift occurs along the direction of the \SNIa{}-only contour degeneracy, the effect on the combined \SNIa{}, CMB, and BAO results are negligible for the DES data set. Furthermore the shift is still negligible in an \SNIa{}-only analysis relative to the parameter uncertainties, 0.07 and 0.35 respectively for $\Omega_{\rm m}$ and $w$.

\subsection{Results on individual CCDs}
\label{ccdeffects}
\begin{figure*}[]
\centering
\includegraphics[width=18cm]{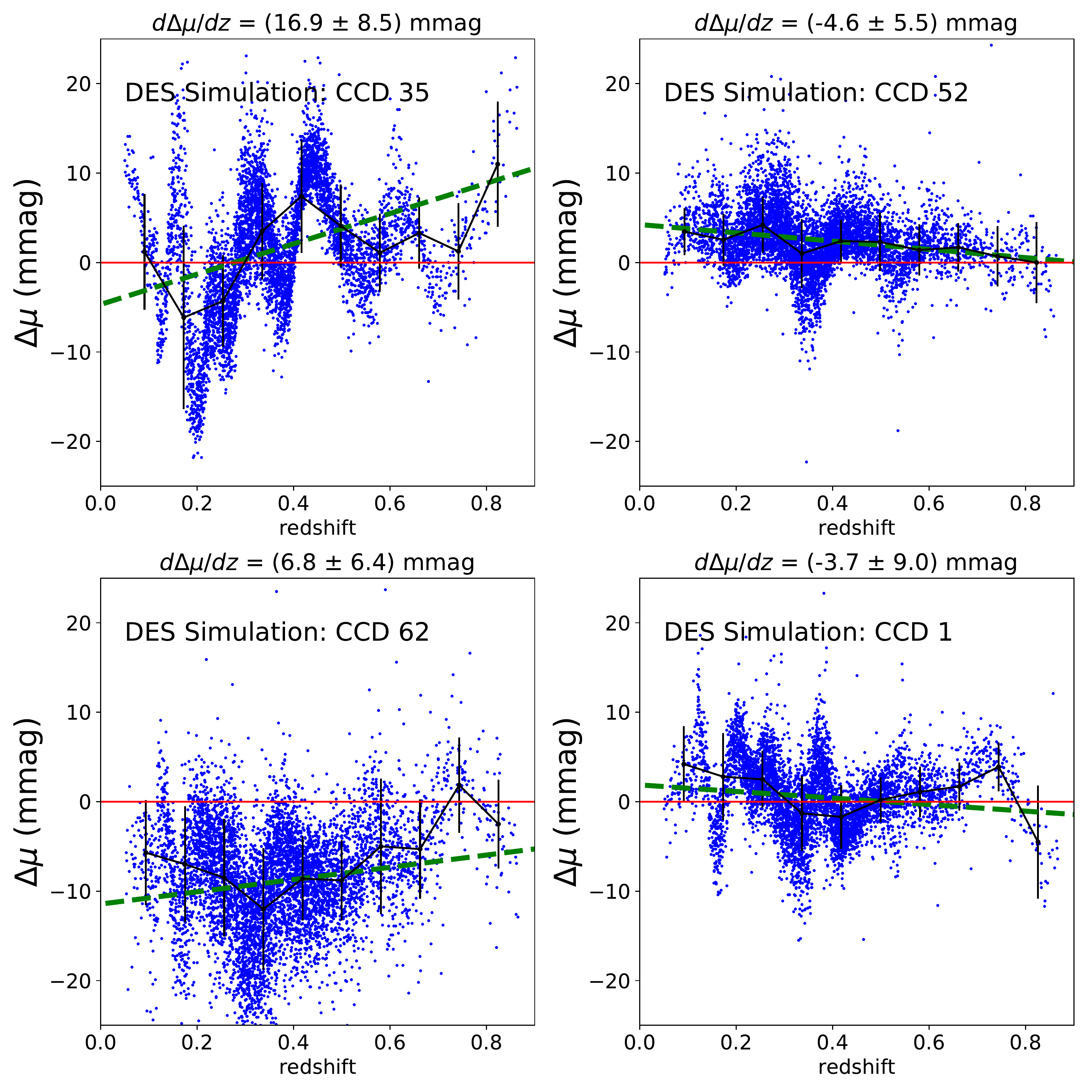}
\caption{For simulated DES-like supernova, $\dmu$ (Equation \ref{eqdmu}) vs.\ redshift for the labeled CCDs.}
\label{CCDPanel}
\end{figure*}

There is no significant trend in $\dmu$ vs redshift as shown in Figs.~\ref{dataPanel} and \ref{simPanel}. However, this is not the case for individual CCDs in the simulated sample. The data sample is too small to get meaningful results for individual CCDs, so we use a simulated sample. Fig.~\ref{CCDPanel} shows $\dmu$ vs redshift for 4 different CCDs at different distances from the center of the focal plane. CCD 35 is near the center, CCD 52 is halfway between the center and the edge, and CCDs 1 and 62 are at the far edge of the focal plane on opposite sides. These 4 CCDs were chosen to sample the radial transmission function variation as shown in L16. These simulations show that some CCDs have a strong trend in $\dmu$ vs. redshift.

The results of fitting these redshift trends for the data, simulation, and subsets of the simulation for each of the chosen CCDs are summarized in the bottom four rows of each panel of Fig.~\ref{SlopeIdeogram}. The scatter among the individual CCD samples is large and in many cases the CCDs are both inconsistent with each other and with zero. The individual corrections show a strong oscillatory behavior with redshift that comes from features of the SED moving into and out of the bandpasses with redshift.

\subsection{Cross Checks}
\label{crosschecks}
Since our redshift cutoff of 0.85 is a function of the DES spectroscopic selection function (D'’Andrea et al 2018 in prep.) and is not related to the chromatic corrections or the supernovae themselves, we have also tested the effect of changing this cutoff to lower redshift. There is no significant change in $\dw$ and $\dOm$ with decreasing redshift cutoff.

To obtain the spectra used in the chromatic corrections, we use the spectral templates from the SALT2 model. This SED model is constructed from spline basis functions and thus may alter some of the SN spectral features.  To check if our chromatic corrections are sensitive to the SALT2 SED representation, we have performed a cross-check based on the spectral time series created by \citet{Hsiao}. \citet{M14} constructed a model from the Hsiao spectral time series using the SALT2 stretch and color law relations while preserving the spectral features. Using this model results in corrections consistent with those based on the SALT2 model spectra: the $\dmu$ (Eq.\ref{eqdmu}) agree to within 0.25 mmag for $-0.3 < c < +0.3$.

\section{CONCLUSION}
\label{conclusion}

In this paper, we presented the first application of the chromatic corrections described in B18 to type Ia supernova cosmology. We applied the corrections to the \samplename{} supernova sample as defined in the \citealt{KEYPAPER} analysis. The effect of the chromatic corrections on distance modulus is not significant in either data or simulation. The $1\sigma$ limit on the median size of the chromatic correction on the single epoch photometry is a less than 2 mmag change in correction over the redshift range from $z = 0$ to $z = 1$. This limit is valid for the \samplename{} sample and is not necessarily valid for other samples, although this has not been tested. 
The application of chromatic corrections, while necessary to achieve the precision photometry in B18, results in a change in  $w$ of $\wcorr{} \pm \wcorrerr{}$ and a change in $\Omega_m$ of $\omcorr{} \pm \omcorrerr{}$,  for the combined \SNIa{}, BAO, and CMB analysis, which is not statistically significant. Examining the effect of the corrections on single CCDs both in $\dmu$ trends vs.\ redshift and cosmological parameters shows that this effect would become significant on a targeted survey where the observations are placed on a single CCD or subset of CCDs. This is assuming that such a survey would use the FGCM calibration of DES and would not recalibrate itself to a reference spectrum based only on the CCDs it used.

\acknowledgments

James Lasker is grateful for support from the ARCS (Acheivement Rewards for College Scholars) foundation and its donors. James Lasker, Rick Kessler, Dan Scolnic, and Josh Frieman are grateful for the support of the University of Chicago Research Computing Center for assistance with the calculations carried out in this work and to the Kavli Institute for Cosmological Physics.

Dan Scolnic is supported by NASA through Hubble Fellowship grant HST-HF2-51383.001 awarded by the Space Telescope Science Institute, which is operated by the Association of Universities for Research in Astronomy, Inc., for NASA, under contract NAS 5-26555.  

Funding for the DES Projects has been provided by the U.S. Department of Energy, the U.S. National Science Foundation, the Ministry of Science and Education of Spain, 
the Science and Technology Facilities Council of the United Kingdom, the Higher Education Funding Council for England, the National Center for Supercomputing 
Applications at the University of Illinois at Urbana-Champaign, the Kavli Institute of Cosmological Physics at the University of Chicago, 
the Center for Cosmology and Astro-Particle Physics at the Ohio State University,
  the Mitchell Institute for Fundamental Physics and Astronomy at Texas A\&M University, Financiadora de Estudos e Projetos, 
  Funda{\c c}{\~a}o Carlos Chagas Filho de Amparo {\`a} Pesquisa do Estado do Rio de Janeiro, Conselho Nacional de Desenvolvimento Cient{\'i}fico e Tecnol{\'o}gico and 
  the Minist{\'e}rio da Ci{\^e}ncia, Tecnologia e Inova{\c c}{\~a}o, the Deutsche Forschungsgemeinschaft and the Collaborating Institutions in the Dark Energy Survey. 

  The Collaborating Institutions are Argonne National Laboratory, the University of California at Santa Cruz, the University of Cambridge, Centro de Investigaciones Energ{\'e}ticas, 
  Medioambientales y Tecnol{\'o}gicas-Madrid, the University of Chicago, University College London, the DES-Brazil Consortium, the University of Edinburgh, 
  the Eidgen{\"o}ssische Technische Hochschule (ETH) Z{\"u}rich, 
  Fermi National Accelerator Laboratory, the University of Illinois at Urbana-Champaign, the Institut de Ci{\`e}ncies de l'Espai (IEEC/CSIC), 
  the Institut de F{\'i}sica d'Altes Energies, Lawrence Berkeley National Laboratory, the Ludwig-Maximilians Universit{\"a}t M{\"u}nchen and the associated Excellence Cluster Universe, 
  the University of Michigan, the National Optical Astronomy Observatory, the University of Nottingham, The Ohio State University, the University of Pennsylvania, the University of Portsmouth, 
  SLAC National Accelerator Laboratory, Stanford University, the University of Sussex, Texas A\&M University, and the OzDES Membership Consortium.

  Based in part on observations at Cerro Tololo Inter-American Observatory, National Optical Astronomy Observatory, which is operated by the Association of 
  Universities for Research in Astronomy (AURA) under a cooperative agreement with the National Science Foundation.

  The DES data management system is supported by the National Science Foundation under Grant Numbers AST-1138766 and AST-1536171.
  The DES participants from Spanish institutions are partially supported by MINECO under grants AYA2015-71825, ESP2015-66861, FPA2015-68048, SEV-2016-0588, SEV-2016-0597, and MDM-2015-0509, 
  some of which include ERDF funds from the European Union. IFAE is partially funded by the CERCA program of the Generalitat de Catalunya.
  Research leading to these results has received funding from the European Research
  Council under the European Union's Seventh Framework Program (FP7/2007-2013) including ERC grant agreements 240672, 291329, and 306478.
We  acknowledge support from the Australian Research Council Centre of Excellence for All-sky Astrophysics (CAASTRO), through project number CE110001020, and the Brazilian Instituto Nacional de Ci\^encia
e Tecnologia (INCT) e-Universe (CNPq grant 465376/2014-2).

This manuscript has been authored by Fermi Research Alliance, LLC under Contract No. DE-AC02-07CH11359 with the U.S. Department of Energy, Office of Science, Office of High Energy Physics. The United States Government retains and the publisher, by accepting the article for publication, acknowledges that the United States Government retains a non-exclusive, paid-up, irrevocable, world-wide license to publish or reproduce the published form of this manuscript, or allow others to do so, for United States Government purposes.
\bibliography{ChCorr}

\end{document}